\documentclass[%
 reprint,
nofootinbib,
showkeys,
 amsmath,amssymb,
 aps,
pra
]{revtex4-2}
\usepackage{appendix}
\usepackage{graphicx}
\usepackage{microtype}
\usepackage{enumerate}
\usepackage{empheq}
\usepackage{tabularx}
\usepackage{blindtext}
\usepackage[T2A]{fontenc}
\usepackage{tikz}

\usepackage{physics}
\usepackage{marvosym}

\usepackage{float}
\usepackage{import} 
\usepackage{dcolumn}
\usepackage{bm}
\usetikzlibrary{patterns}
\usepackage{dsfont}
\usepackage{commath}
\usepackage[unicode=true,pdfusetitle,
bookmarks=true,bookmarksnumbered=false,bookmarksopen=false,
breaklinks=false,pdfborder={0 0 1},backref=false,colorlinks=false]
{hyperref}
\usepackage{ulem}

\hypersetup{
	colorlinks,linkcolor=blue,citecolor=blue,urlcolor=blue}

\begin{document}

\preprint{BJP/123-QED}

\title{On the perturbed harmonic oscillator and celestial mechanics}


\author{J. Oliveira-Cony}

 \email{joaooctavio8@gmail.com}
\affiliation{%
 Universidade Federal do Rio de Janeiro - UFRJ\\
 Instituto de Física 
}%

\author{C. Farina}

 \email{carfarina@gmail.com}
\affiliation{%
 Universidade Federal do Rio de Janeiro - UFRJ\\
 Instituto de Física 
}%

\date{\today}

\begin{abstract}

We study the influence of perturbations in the three dimensional isotropic harmonic oscillator problem considering different perturbing force laws and apply our results in the context of celestial mechanics, particularly in the movement of stars in stellar clusters. We use a method based on the Runge-Lenz tensor, so that our results are valid for any eccentricity of the unperturbed orbits of the oscillator. To establish basic concepts, we start by considering two cases, namely: a Larmor and a keplerian perturbation; and show that, in both cases, the perturbed orbits will precess. After that, we consider the more general problem of a central perturbation with any power-law dependence, that also only causes precession. Then, we consider precessionless perturbations caused by an Euler force and by the non-central dragging forces of the form  
$\boldsymbol{\delta F}=-\gamma_nv^{n-1}\boldsymbol{v}$, where $\boldsymbol{v}$ is the velocity of the particle and $\gamma_n\geq0$. 
We demonstrate that, in the case of a linear drag $(n=1)$, the orbits eccentricities remains constant. In contrast to what occurs in  the well-known Kepler problem, 
for $n>1$ the orbit becomes increasingly eccentric. In the case $n=-3$, where the force is interpreted as a Chandrasekhar friction, we show that the eccentricity diminishes over time. We finish this work by making a few comments about the relevance of the main results.
\end{abstract}

\maketitle


\section{Introduction}
\label{Intro}

\label{Intro}

In 1873, J. Bertrand proved an important theorem which states that the only central forces for which all bounded orbits are closed are the inverse square force and the harmonic oscillator one  \cite{bertrand1873}. For other demonstrations, considerations and comments on the theorem, see \cite{Brown1978,Tikochinsky1988,Zarmi2002,arnold2007mathematical,Santos2009,Chin2015,lemos2018}.

The study of orbits in the context of planetary motion dates from prior to Ptolomaeus \textit{Almagest} \cite{Ptolemy2020}, and has its epitome in the works of Kepler on the XVII century. In fact, in his famous {\it Astronomia Nova} \cite{kepler2015} published in 1609, he presented, among many other results, what is known today as his first law (planets move in elliptic orbits with one of their foci located at the center of force)\footnote{For the harmonic oscillator the orbits are also ellipses, but with the center of the ellipse located at the center of force.}
and also his second law (the areas swept by the position vector of a planet for equal time intervals are always the same).\footnote{What is known as Kepler's third law - the ratio between the square of the period and the major semiaxis to the power three is the same for all planets - appeared for the first time only in 1619 in Kepler's {\it Harmonices Mundi}, see Ref. \cite{Kepler1952}.}
Kepler's laws, together with Bertrand's theorem, form the basis of the study of closed orbits.

Given Bertrand's theorem, it is natural to consider Kepler's problem and the harmonic oscillator in similar scenarios. While Kepler's potential is the most common one in astronomical applications, stars in stellar clusters are subjected, in a good approximation, to a harmonic potential since, for the length scales involved, clusters can be considered as spheres with uniform densities. In this way, this potential plays a fundamental role in understanding how stars move in such clusters. The same behaviour is expected for galaxies in galaxy clusters.

However, the harmonic potential describes a very idealized situation. In more realistic cases, we must take into account many types of perturbations. A good strategy to analyze the influence of perturbations on the unperturbed elliptical orbits of a three dimensional isotropic harmonic oscillator (3DIHO) is to study the orbit precession caused by such perturbations. In the cases where the perturbations do not cause any precession, it is interesting to study the changes caused in the orbits' eccentricity 
\cite{Sivardire1984,Farina1988,Tort1989}. Although there is an extensive literature on the study of the Kepler problem under different kinds of perturbations, the same can not be said for the 3DIHO, whose number of works is very scarce in the literature compared to the Kepler problem. One of our main goals is to fill this gap.

%



The method to be employed here is based on a conserved quantity on the unperturbed 3DIHO, the Runge-Lenz tensor \cite{Sivardire1984}. 
%
%
This quantity was first described as a vector for the Kepler problem and then was generalized for arbitrary central forces by Fradkin \cite{Fradkin1965,Fradkin1967} and, independently, by A. Peres \cite{Peres1979} - for the equivalence of their demonstrations, see \cite{Yoshida1987}. For the Kepler problem, the modulus of the Runge-Lenz vector is proportional to the eccentricity of the elliptical orbit, while in the harmonic oscillator problem this invariant takes the form of a second-order tensor. This is directly related to the fact that the center of force is at the center of the ellipse, which explains the existence  of two axes of symmetry passing through the center of force, in contrast with a unique symmetry axis exhibited by the keplerian orbits.

The history of the use of this quantity in the Kepler (or Coulomb) problem is more convoluted \cite{Goldstein1975,Goldstein1976}. In 1926, W. Pauli used this vector to calculate the energy spectrum of the hydrogen atom \cite{Pauli1926}, and commented that the invariant had been used previously by W. Lenz, who, in 1924, used the vector in the context of the old quantum theory \cite{Lenz1924}. Lenz cites in his work C. Runge's vector analysis book of 1919 \cite{runge1923vector}, and then the name Runge-Lenz was coined. As P.S. Laplace discussed in detail this quantity in his \textit{Traité de Mécanique Celeste} \cite{Laplace1799} of 1799, the vector was repopularised later as Laplace-Runge-Lenz (LRL) vector. But Laplace was not the first one to describe this quantity, and the merit of its discovery is due to J. Hermann in 1710, who utilized this concept to find the orbit equation  the  Kepler problem \cite{Hermann1710}. In 1712, Johann Bernoulli generalized the result for arbitrary orbit orientations. In this sense, a more befitting name for this quantity would be Hermann-Bernoulli-Laplace-Runge-Lenz-Pauli invariant. For the sake of brevity, we will call it the $\mathds{A}-$invariant. This quantity continued to be used in important contexts principally in mathematical physics \cite{BenYaacov2010, Ballesteros2011, Hakobyan2015,Pain2022, Efimov2023}, but also in classical mechanics - for example, in the study of solar sails \cite{Garrido2023} and extremal and supermassive black holes \cite{Neeling2024,Fang2020} -, and in quantum theory \cite{Jauch1940,Heintz1974,Farina2011,Glikov2013,Silenko2013, Hey2015}. In particular, it was recently showed that, in the Kepler problem, there are limitations for when the modulus of the Runge-Lenz vector is a measure for the eccentricity, where the full vector character appears \cite{Will2019}.

In this paper, we study systematically different kinds of perturbing forces on the 3DIHO using the properties of the second rank tensor $\mathds{A}$, whose eigenvectors give the directions of the two symmetry axes of the unperturbed orbits, while their eigenvalues are directly related to the eccentricity  of these orbits. We then consider possible applications in the context of celestial mechanics.

This paper is organized as follows. In Sec.~\ref{sec2}, 
we describe the method to be used to analyze the main characteristics of the perturbed orbits, namely, their velocities of angular precession and their eccentricities. 
In Sec.~\ref{sec3} we discuss many kinds of perturbations that cause precession. As a warming up problem, we start  with the Larmor force and then we proceed with a keplerian perturbation. We finish this section generalizing our calculations to a perturbing central potential with any power law.  In Sec.~\ref{sec4} we consider forces that do not cause precession, but instead change the lengths of both semiaxes of the perturbed orbits. We solve the simple case of the Euler force and then we consider the interesting Chandrasekhar friction, an important effective force in the context of stars in stellar clusters. Finally, we solve the general case of air-resistance-like perturbations $\boldsymbol{\delta F}_n=-\gamma_nv^{n-1}\boldsymbol{v}$, where $\boldsymbol{v}$ is the velocity of the oscillator and $\gamma_n>0$. Particularly, we demonstrate that, in the linear drag case ($n=1$), the eccentricity of the perturbed orbit remains constant as it curls up before collapsing to the center of force. We trace our conclusions in Sec. \ref{conclusion}.


\section{Perturbed Orbits of the Harmonic Oscillator}
\label{sec2}

Let us start by considering the unperturbed 3DIHO, that is, a particle of mass $m$ subjected to a central force of the form $\boldsymbol{F}_{\mathrm{HO}}=-m\omega^2\boldsymbol{r}$, 
where $\boldsymbol{r}$ is the position of the particle relative to the center of force, and $\omega>0$ is its natural frequency. The corresponding orbits are in general ellipses with centers at the center of force. In what follows, we will denote by $a$ the major semiaxis and by $b$ the minor one. 

For the harmonic oscillator, the generalized $\mathds{A}-$invariant is given by \cite{Fradkin1965}
\begin{equation}
    \mathds{A}=\dfrac{1}{2m}\boldsymbol{p}\otimes \boldsymbol{p}+\dfrac{m\omega^2}{2}\boldsymbol{r}\otimes\boldsymbol{r}\,,
\end{equation}

\noindent where $\boldsymbol{p}=m\dfrac{d\boldsymbol{r}}{dt}$. For an alternative definition, see Appendix \ref{appB}. This quantity is easy to interpret when we write the matrix of $\mathds{A}$ in the basis of the direction of the axis of oscillation. If the motion of the oscillator is such that, in a given Cartesian coordinate system, $x=a\cos(\omega t)$ and $y=b\sin(\omega t)$, then in the basis $\mathcal{C}=\{\hat{x},\hat{y},\hat{z}\}$, it takes the matrix form
\begin{align}
    [[\mathds{A}]]_{\mathcal{C}}=\begin{bmatrix}
        \dfrac{m\omega^2 a^2}{2} & 0 & 0\\
        0 & \dfrac{m\omega^2 b^2}{2} & 0\\
        0  & 0 & 0
    \end{bmatrix}
    =:\begin{bmatrix}
        A & 0\\
        0 & 0
    \end{bmatrix}\,.
    \label{Matrix}
\end{align}
The eigenvalues $(\lambda_x,\lambda_y,\lambda_z)=m\omega^2/2(a^2,b^2,0)$ of $\mathds{A}$ are, then, the energies associated with the directions of the eigenvectors, and so $\mathds{A}$ is conserved. As there is no movement on the $\mathcal{Z}-$axis, it was convenient to define in the previous equation the block matrix $A$. Since, by assumption,  $a>b$, the eccentricity of the orbit is written as
\begin{equation}
    \epsilon:=\sqrt{1-\dfrac{b^2}{a^2}}=\sqrt{1-\dfrac{\lambda_y^2}{\lambda_x^2}}\,.\label{Eccentricity}
\end{equation}

Now, let us include a perturbing force in our discussion so that the total force on the 3DIHO is $\boldsymbol{F}=\boldsymbol{F}_{\mathrm{HO}}+\boldsymbol{\delta F}$. Computing the time derivative and using Newton's second law, we obtain
%
%
\begin{eqnarray}
   \dfrac{d\mathds{A}}{dt} &=& \dfrac{1}{2m}\dfrac{d}{dt}\boldsymbol{p}\otimes \boldsymbol{p}+\dfrac{m\omega^2}{2}\dfrac{d}{dt}\boldsymbol{r}\otimes\boldsymbol{r}\cr\cr
   &=&\dfrac{1}{2}\left(\boldsymbol{F}\otimes \boldsymbol{v}+\boldsymbol{v}\otimes \boldsymbol{F}\right)+\dfrac{m\omega^2}{2}\left(\boldsymbol{v}\otimes\boldsymbol{r}+\boldsymbol{r}\otimes\boldsymbol{v}\right)\,.
\end{eqnarray}
Since $ \boldsymbol{F}=-m\omega^2\boldsymbol{r}+\boldsymbol{\delta F}$, we get
\begin{equation}
    \dfrac{d\mathds{A}}{dt}=\dfrac{1}{2}\left(\boldsymbol{\delta F}\otimes \boldsymbol{v}+ \boldsymbol{v}\otimes \boldsymbol{\delta F}\right)\,.
    \label{defDerivative}
\end{equation}
Taking $\boldsymbol{\delta F}=0$, we immediately see that   $\mathds{A}$ is indeed a constant of motion of the unperturbed 3DIHO. Note that, in general, for the block matrix $A$ we have
\begin{equation}
    \dfrac{dA}{dt}=\begin{bmatrix}
    f_{xx} & 0\\
        0 & f_{yy} 
    \end{bmatrix}+f_{xy}\begin{bmatrix}
        0 & 1 \\
        1 & 0 \\
    \end{bmatrix}\,.
\end{equation}
The term $f_{xy}$ is associated with the precession of the orbit, while $f_{xx},f_{yy}$ are related, respectively, with the length variation of the largest and smallest semiaxis. Particularly, if $f_{xx} = f_{yy}=0$, then $f_{xy}$ is proportional to the angular velocity of precession.\footnote{If $f_{xy}=0$, as it will be the case in the next section, the rate of change of the length of the semiaxis, $a$ and $b$, are directly related to $f_{xx}$ and $f_{yy}$. If $f_{xx},f_{yy},f_{xy}\neq0$, then there is precession and there is variation on the length of the semiaxis, but because of the non-commutative relation of matrices, the coefficients are mixed in the velocities descriptions. These calculations are possible and straightforward, but are less clear, so we opted to not consider them in this paper.}
In order to leave this interpretation more evident, we consider an infinitesimal pure rotation of the $A$ matrix by an angle $d\theta$, so that
\begin{equation}
    A+dA=dR\,A\,dR^{-1}
\end{equation}
where
\begin{equation}
    dR=\begin{bmatrix}
        1 & 0\\
        0 & 1
        \end{bmatrix}+\begin{bmatrix}
        0 & -d\theta\\
        d\theta & 0
    \end{bmatrix}\,.
\end{equation}
In this way, considering Eq. \ref{Matrix}, we may write
\begin{equation}
   \dfrac{dA}{dt}=\dfrac{m\omega^2}{2}(a^2-b^2)\Omega\begin{bmatrix}
        0 & 1\\
        1 & 0
    \end{bmatrix}\,.
    \label{press}
\end{equation}
where $\Omega=d\theta/dt$ is identified as the velocity of angular precession. In order to obtain the velocity of precession in first order of perturbation, the left hand side of the previous equation is substituted by its time average over the unperturbed orbit, given by
%
%
\begin{align}
    \left\langle\dfrac{dA}{dt}\right\rangle_{ij} =\dfrac{\omega}{2\pi}\int_{-\pi/\omega}^{\pi/\omega}(F_iv_j+F_jv_i)\,dt\quad(i,j=1,2)\,.
    \label{Adetalhado}
\end{align}

%
Whenever $\langle dA/dt\rangle_{ii}=0$, the substitution of the previous equation into the left hand side of Eq. \ref{press} will lead to the desired velocity of precession, as we will show explicitly in the next section.


\section{Precessing Perturbations}
\label{sec3}

In this section, we treat perturbing forces that only causes precession. We begin with the simple example of a perturbing Larmor force on a charged isotropic oscillator. Then, we discuss central perturbations, starting with a keplerian one, and after that we generalize our results considering central perturbations with any power law.


\subsection{Larmor force}

As a classical example of force that causes precession, we consider a Larmor-type force, 
$\boldsymbol{\delta F}=-q\boldsymbol{v}\times\boldsymbol{B}$ where $\boldsymbol{B}=B\hat{z}$ is a constant and uniform magnetic field and $q$ is the charge of the harmonic oscillator. Using Eq. \ref{Adetalhado}, we get
\begin{equation}
    \left\langle\dfrac{dA}{dt}\right\rangle = \dfrac{qB}{2}\begin{bmatrix}
        \langle v_xv_y \rangle &\langle v_y^2-v_x^2\rangle\\
        \langle v_y^2-v_x^2\rangle & \langle v_xv_y \rangle
    \end{bmatrix}\,.
\end{equation}
Since $\langle \cos(\omega t) \sin(\omega t)\rangle=0$ and $\langle \cos^2(\omega t)\rangle=\langle \sin^2(\omega t)\rangle=\frac{1}{2}$, the previous equation takes the form
\begin{equation}
    \left\langle\dfrac{dA}{dt}\right\rangle = \dfrac{qB\omega^2}{4}(b^2-a^2)\begin{bmatrix}
        0 & 1\\
        1 & 0
    \end{bmatrix}\,.
\end{equation}
Comparing this result with Eq. \ref{press}, we readily identify  $\Omega=-qB/2m$, which is the expected Larmor precession for a charged oscillator with charge $q$ immersed in a magnetic field $\boldsymbol{B}$.

\begin{figure}[h!]
    \centering
    \includegraphics[width=1\linewidth]{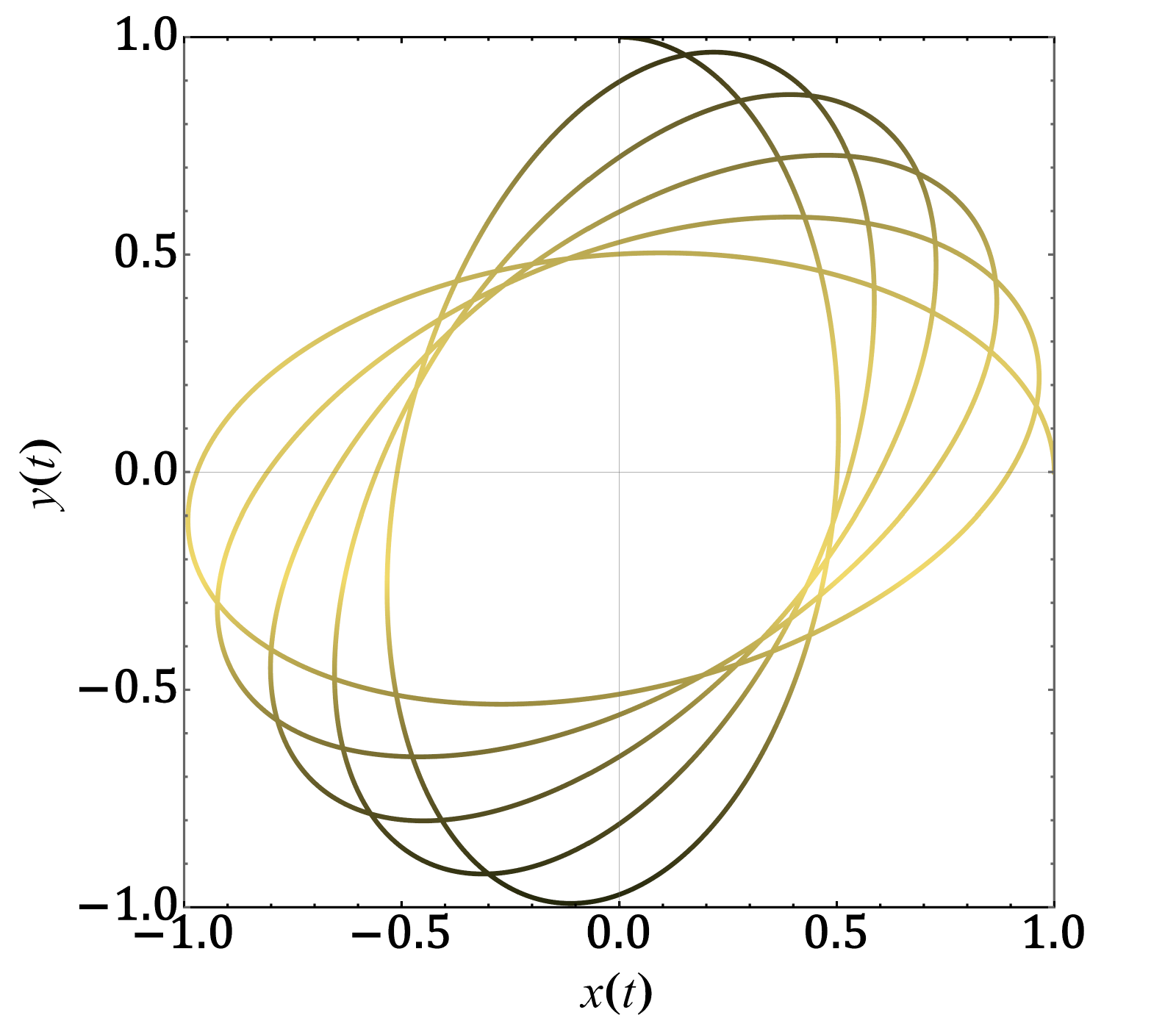}
    \caption{Orbit of a charged 3DIHO submitted to a perturbation given by a constant and uniform magnetic field perpendicular to the plane of the orbit. In this graph we used $a=1$, $b=0.5$, and $\Omega=0.05$ in arbitrary units.
    }
    \label{Precc}
\end{figure}

 In Fig. \ref{Precc}, we plot a possible perturbing orbit for this case assuming arbitrary  values for $q$, $m$, $B$ and the orbit parameters. In general, the perturbed orbit is not closed anymore, except for  when $\Omega\tau$ and $2\pi$ are commensurable, $\tau$ being the period of the unperturbed orbit.


\subsection{Keplerian force}

In a stellar cluster a star is subject, in a first approximation (to be improved in a moment), to an isotropic harmonic potential, since in the relevant length scales for a given stellar orbit, the cluster can be considered as a sphere with a uniform mass distribution  described by a volumetric density of mass $\rho$. Note that there is no problem in considering  a star in the cluster as a particle, since the length scale of a star is negligible when compared to the dimension of a stellar cluster.\footnote{A typical stellar cluster may contain from dozens of thousands to millions of stars and a typical distance between two stars in a cluster is approximately one light-year.}
 Using Gauss's law, the force on a given star of mass $m$ inside the cluster is given by
\begin{equation}
\boldsymbol{F}_{\mathrm{HO}}=-m\dfrac{4\pi G\rho}{3}\boldsymbol{r}=:-m\omega^2\boldsymbol{r}\,.    
\label{Virial}
\end{equation}
where $G$ is the universal gravitational constant.

\begin{figure}[h!]
    \centering
    \includegraphics[width=0.5\linewidth]{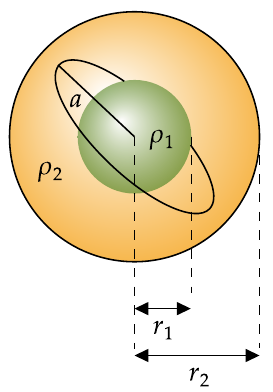}
    \caption{Model of a cluster consisting of a sphere of radius $r_1$ and a concentric spherical shell of radii $r_1$ and $r_2$. The mass densities of the sphere and the shell are  $\rho_1$ and $\rho_2 < \rho_1$, respectively. The prescribed (unperturbed) ellipse of a given star has semiaxis of length $a$ and $b$ such that $r_1 <b<a < r_2$.}
    \label{Shell}
\end{figure}

The previous approximation is clearly a great idealization
of the structure of a stellar cluster. Since there is a concentration of stars in the central region of the cluster, a more realistic model for it consists of a spherical kernel of radius $r_1$ and a concentric spherical shell of internal radius $r_1$ and external one $r_2$. For simplicity, let us suppose that
 mass densities of the sphere and the shell are uniform and given by $\rho_1$ and $\rho_2$, respectively, with $\rho_1 > \rho_2$. Dee 
 Fig. \ref{Shell} for an illustration of this simple model.
 

Suppose a given star in the cluster describes an (yet unperturbed) orbit whose semiaxes satisfy $r_1<b<a<r_2$, as shown in Fig.~\ref{Shell}. In this case, 
%
%
the net force on the star can be separated into two contributions: one coming from a uniform mass distribution of density $\rho_2$ in the whole cluster and a keplerian gravitational contribution caused by a spherical kernel of radius $r_1$ and density $\rho_1 - \rho_2$.
%
%
The first contribution gives rise to the harmonic force $\boldsymbol{F}_{\mathrm{HO}}=-m\omega^2\boldsymbol{r}$, where $\omega$ depends only on $\rho_2$, while the second one can be viewed as a perturbation if we assume that  
$\vert \boldsymbol{\delta F}\vert/\vert{\boldsymbol F}_{\mathrm{HO}}\vert \ll 1$ for all points of the stellar orbit. From Gauss law, we ready obtain\footnote{To be considered a perturbation, it is necessary that $kb^{-2}\ll m\omega^2 b$ or, analogously, $(\rho_1-\rho_2)/\rho_2\ll 4\pi b^3/3r_1^3$.}
\begin{equation}
    \boldsymbol{\delta F}=-\dfrac{Gmr_1^3(\rho_1-\rho_2)}{r^2}\hat{\boldsymbol{r}}=:-kr^{-3}\boldsymbol{r}\, .
\end{equation}
Calculating the time derivative of $A$, we have
\begin{equation}
    \dfrac{dA}{dt}=-\dfrac{k}{2r^3}\begin{bmatrix}
        2v_x x & v_xy+v_yx\\
        v_xy+v_yx & 2v_yy
    \end{bmatrix}\,.
\end{equation}
As the diagonal elements are odd functions of $t$, their time averages vanish, so that
\begin{equation}
    \left\langle\dfrac{dA}{dt}\right\rangle=-\dfrac{kab\omega}{2}\left\langle\dfrac{\cos^2(\omega t)-\sin^2(\omega t)}{r^3}\right\rangle\begin{bmatrix}
        0 & 1\\
        1 & 0
    \end{bmatrix}\,.
\end{equation}

In order to calculate the previous time average, note initially that
%
%
\begin{align}
r^3&=(a^2\cos^2(\omega t)+b^2\sin^2(\omega t))^{3/2}\nonumber\\
&=a^{3}\left(1-\sin^2(\omega t)+\dfrac{b^2}{a^2}\sin^2(\omega t)\right)^{3/2}\nonumber\\
&=a^{3}(1-\epsilon^2\sin^2(\omega t))^{3/2}\,.
\end{align}
Changing the variable of  integration for $\xi=\omega t$ and considering the parity and periodicity of the functions, we get
\begin{equation}
\left\langle\dfrac{\cos^2(\omega t)-\sin^2(\omega t)}{r^3}\right\rangle
=
\dfrac{2a^{-3}}{\pi}\!\!\int_0^{\pi/2}\!\!\dfrac{\cos^2(\xi)-\sin^2(\xi)}{(1-\epsilon^2\sin^2(\xi))^{3/2}}d\xi\,.
\label{AKepler}
\end{equation}
This integral can be written in terms of the so-called hypergeometric function $_{2}F_1$ (see Appendix \ref{appA}):
\begin{align}
    \int_0^{\pi/2}\cos^2(\xi)(1-\epsilon^2\sin^2(\xi))^{-3/2}\,d\xi&=\dfrac{\pi}{4}{_2}F_1(3/2,1/2,2,\epsilon^2)\,,\\
    \int_0^{\pi/2}\sin^2(\xi)(1-\epsilon^2\sin^2(\xi))^{-3/2}\,d\xi&=\dfrac{\pi}{4}{_2}F_1(3/2,3/2,2,\epsilon^2)\,.
\end{align}
Since ${_2}F_1(3/2,3/2,2,\epsilon^2)\geq {_2}F_1(3/2,1/2,2,\epsilon^2)$ on the interval $\epsilon\in [0,1)$, then the time average is  negative, and consequently the precession will be positive, that is, the orbit precesses so that there will be an advance of its pericenter (or simply anti-clockwise, if we suppose the angular momentum of the star points outwards the page). Substituting this result in Eq.(\ref{AKepler}), we obtain
\begin{align}
\Omega&=\dfrac{k}{m\omega^2}\dfrac{a^{-3}}{(a^2-b^2)}\bigg({_2}F_1(3/2,3/2,2,\epsilon^2)\; +\nonumber\\&\hspace{4cm}-{_2}F_1(3/2,1/2,2,\epsilon^2)\bigg)\nonumber\\
&=\dfrac{k'}{\epsilon^{2}}\bigg({_2}F_1(3/2,3/2,2,\epsilon^2)\; -{_2}F_1(3/2,1/2,2,\epsilon^2)\bigg)\,,
\end{align}

\noindent where $k'=k/m\omega^2a^5$.

This model can be readily improved in order to describe more realistic situations by just considering an arbitrary number of spherical shells.

\subsection{Generic $r-$power central force}

The previous calculation for the keplerian perturbing force can be generalized for any power law $\boldsymbol{\delta F}=-\sum_{n}k_nr^{n-1}\boldsymbol{r}$ (but the following results are also valid if $n\notin \mathds{Z}$). 
In this case, it is immediate to see that the corresponding velocity of angular precession is given by $\Omega=\sum_n\Omega_n$ with
\begin{align}
\Omega_n=\dfrac{k_n'}{\epsilon^2}\bigg(&{_2}F_1((n-1)/2,3/2,2,\epsilon^2)+\nonumber\\&-{_2}F_1((n-1)/2,1/2,2,\epsilon^2)\bigg)\,,
\label{GeneralPrecession}
\end{align}
where $k_n'={k_na^{n-3}}/{m\omega^2}$.

With this result, we have solved the perturbation for any $r-$dependent function that has a Taylor expansion. This calculation can be useful, for instance, when we want to improve the model of a sphere and a spherical shell we discussed previously. 
The above calculation can also be useful if the kernel of the cluster has not a spherical symmetry, but still has an axial symmetry, as long as in this case we consider equatorial orbits.

\begin{figure}[H]
    \centering
    \includegraphics[width=1\linewidth]{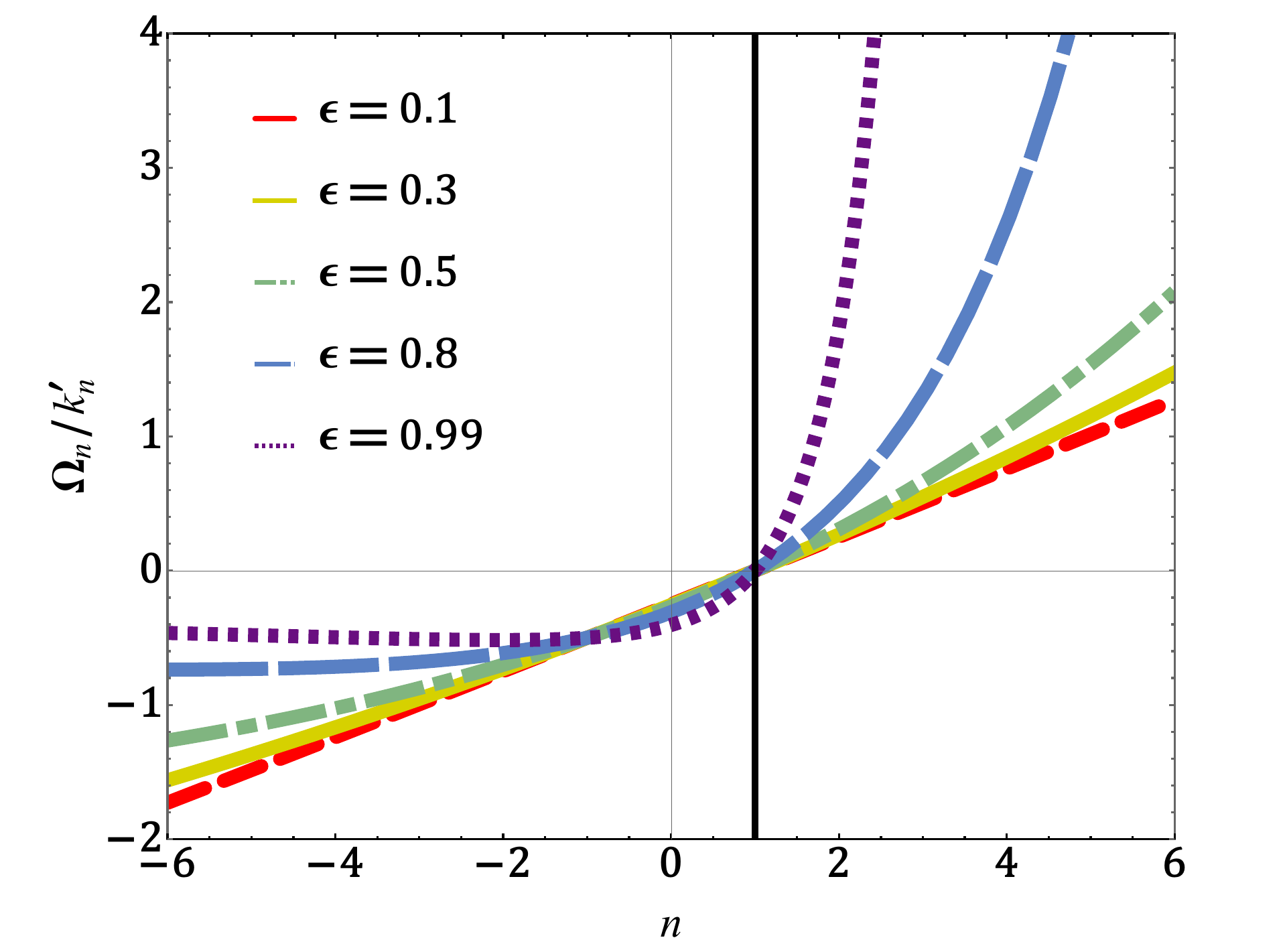}
    \caption{Plot of $\Omega_n/k_n'$ as a function of $n$ for fixed values of the eccentricity. Note that nothing imposes $n$ to be an integer. A black line at $n=1$ separates the anti-clockwise and clockwise precessions.}
    \label{OmegaN}
\end{figure}

Note that, for $k_j'=k_1'\delta_{1j}$, $\Omega=0$, as expected since the perturbing force law in this case is also a harmonic one, $\boldsymbol{\delta F}=-k_1\boldsymbol{r}$. Suppose now that $k_j'=k_n'\delta_{nj}$ with arbitrary $n$. In this case, it can be shown that the value $n=1$ marks a separation of the clockwise and anticlockwise precessions: for $n>1$, the relation of the hypergeometric functions changes, and the orbit precesses clockwise. 
In Fig.~\ref{OmegaN},  we plot the behaviour of $\Omega_n$ as a function of $n$ for different values of the eccentricity. Note that $\Omega_1 =0$ for all eccentricities.

\section{Precessionless Perturbations}
\label{sec4}

In this section, we shall be concerned only with perturbing forces that do not cause precession. We start by considering an Euler type force, and then proceed to analyse the so-called Chandrasekhar friction \cite{Chandrasekhar1943a}, a dissipative force relevant, among other things, in the calculation of the rate of escape of stars from clusters \cite{Chandrasekhar1943b,Chandrasekhar1943c}. Finally, we discuss the more general case of a dissipative force of the form 
$\boldsymbol{\delta F}=-\gamma_nv^{n-1}\boldsymbol{v}$, for arbitrary $n$.

\subsection{Euler Force}

As a simple example, let us  consider an Euler-type force $\boldsymbol{\delta F}=-m\boldsymbol{\alpha}\times \boldsymbol{r}$, where $\boldsymbol{\alpha}=\alpha\hat{z}$ is a constant vector. This kind of force appears, for instance, when we change from an inertial frame to a non-inertial one that whirls anti-clockwise around the $\mathcal{Z}$-axis with constant angular acceleration $\alpha$\footnote{This change of reference frames generates also a centrifugal force and a Coriolis one and both of them only cause precession.}. Plugging this force into Eq. \ref{defDerivative}, we gets
\begin{equation}
    \dfrac{dA}{dt}=-m\alpha \begin{bmatrix}
        2v_xy & v_xx+v_yy\\
        v_xx+v_yy & 2v_yx
    \end{bmatrix}\,.
\end{equation}

Taking the average over one period, we obtain
\begin{equation}
    \left\langle\dfrac{dA}{dt}\right\rangle=\dfrac{m\omega\alpha ab}{2\pi} \begin{bmatrix}
        1 & 0 \\
        0 & -1
    \end{bmatrix}\,.
\end{equation}

\begin{figure}[H]
    \centering
    \includegraphics[width=1\linewidth]{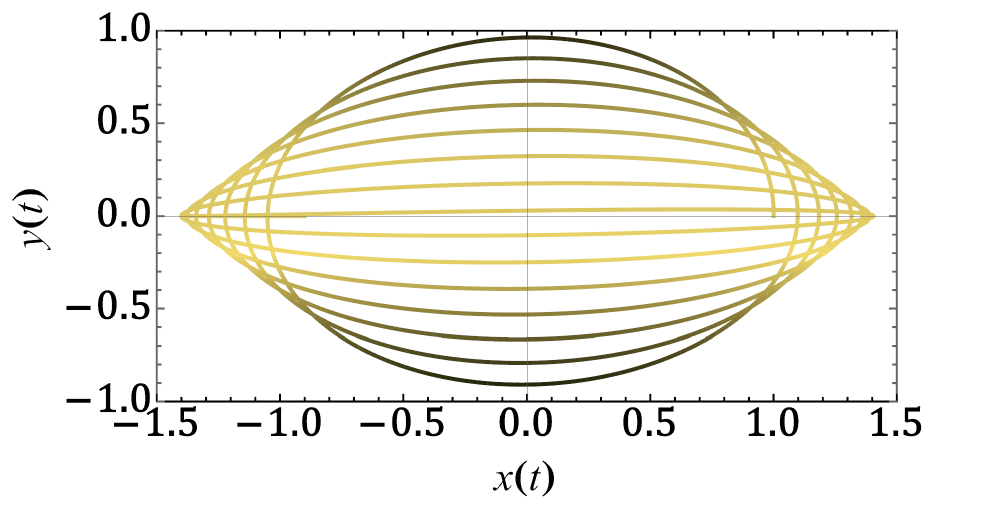}
    \caption{Orbit of the 3DIHO subjected to a perturbation of the form $\boldsymbol{\delta F}=-m\boldsymbol{\alpha}\times \boldsymbol{r}$, with $\alpha=\omega^2/2$ for 16 revolutions. The initial semiaxes are defined as $a=1$ and $b=0.5$ in arbitrary units. Note that there is no precession, but the perturbing orbit becomes more eccentric over time. }
    \label{Euler}
\end{figure}

Together with Eqs. \ref{Matrix} and \ref{Eccentricity}, this result implies that the ellipse will become more eccentric, with a dilation of its largest semiaxis and a contraction of the smallest one (see Fig. \ref{Euler}).

\subsection{Chandrasekhar friction}

In the astronomy domain, one of the main applications of velocity-dependent perturbations is the so-called Chandrasekhar friction (CF) - also called dynamical friction or gravitational drag -, which is the effective force that moving bodies, such as stars or even galaxies, suffer due to the gravitational interaction with the surrounding matter. This force was first described in 1943 by S. Chandrasekhar in the context of star clusters when considering the rate of escape of stars \cite{Chandrasekhar1943a,Chandrasekhar1943b,Chandrasekhar1943c}. For high velocities, a good approximation of this force is given by $\boldsymbol{\delta F}=-Cv^{-3}\boldsymbol{v}$, where $C=\kappa G^2M^2\rho$, $M$ being the mass of the star, $\rho$ the volumetric mass density of the cluster and $\kappa$ a dimensionless constant that depends on the velocity of the surrounding objects \cite{carroll2013introduction}. Note that as $C\propto M^2$, the CF is more expressive in more massive bodies.

The CF can be very important in the evolution of star clusters, since it may explain why massive stars in the cluster tend to concentrate near its center. This fact may give rise to the runaway collision mechanism by which massive objects may be formed in some young and dense star clusters \cite{Zwart2006}. In this sense, analyse the influence of CF on the orbit of a star in a star cluster may be useful to study this kind of process.

Inserting the above mentioned force into the expression for the time derivative of $A$, we obtain
\begin{equation}
\left\langle \dfrac{dA}{dt}\right\rangle=-mC\begin{bmatrix}
    \langle v^{-3}v_x^2 \rangle & 0\\
    0 & \langle v^{-3}v_y^2 \rangle
\end{bmatrix}\,.
\end{equation}
The calculations of the time averages present in the previous equation are very similar to those we made in the previous section. For $\langle v^{-3}v_y^2\rangle$, we get
\begin{align}
    &\dfrac{b^2\omega^2}{2\pi}\int_{-\pi}^{\pi}\sin^2(\xi)[{a^2\omega^2\sin^2(\xi)+b^2\omega^2\cos^2(\xi)]^{-3/2}}\,d\xi \nonumber\\ =&\dfrac{2b^2\omega^2(a\omega)^{-3}}{\pi}\int_0^{\pi/2}\sin^2(\xi)\left[1-\epsilon^2\cos^2(\xi)\right]^{-3/2}\,d\xi \nonumber\\
    =&\dfrac{b^2\omega^2(a\omega)^{-3}}{2}{_2}F_1(3/2,1/2,2,\epsilon^2)\, ,
\end{align}

\noindent and an analogous calculation for  $\langle v^{-3}v_x^2 \rangle$ yields
\begin{equation}
    \langle v^{-3}v_x^2 \rangle=\dfrac{a^2\omega^2 (a\omega)^{-3}}{2} {{_2}F_1(3/2,3/2,2;\epsilon^2)}\,.
\end{equation}

\begin{figure}[ht]
    \centering
    \includegraphics[width=1\linewidth]{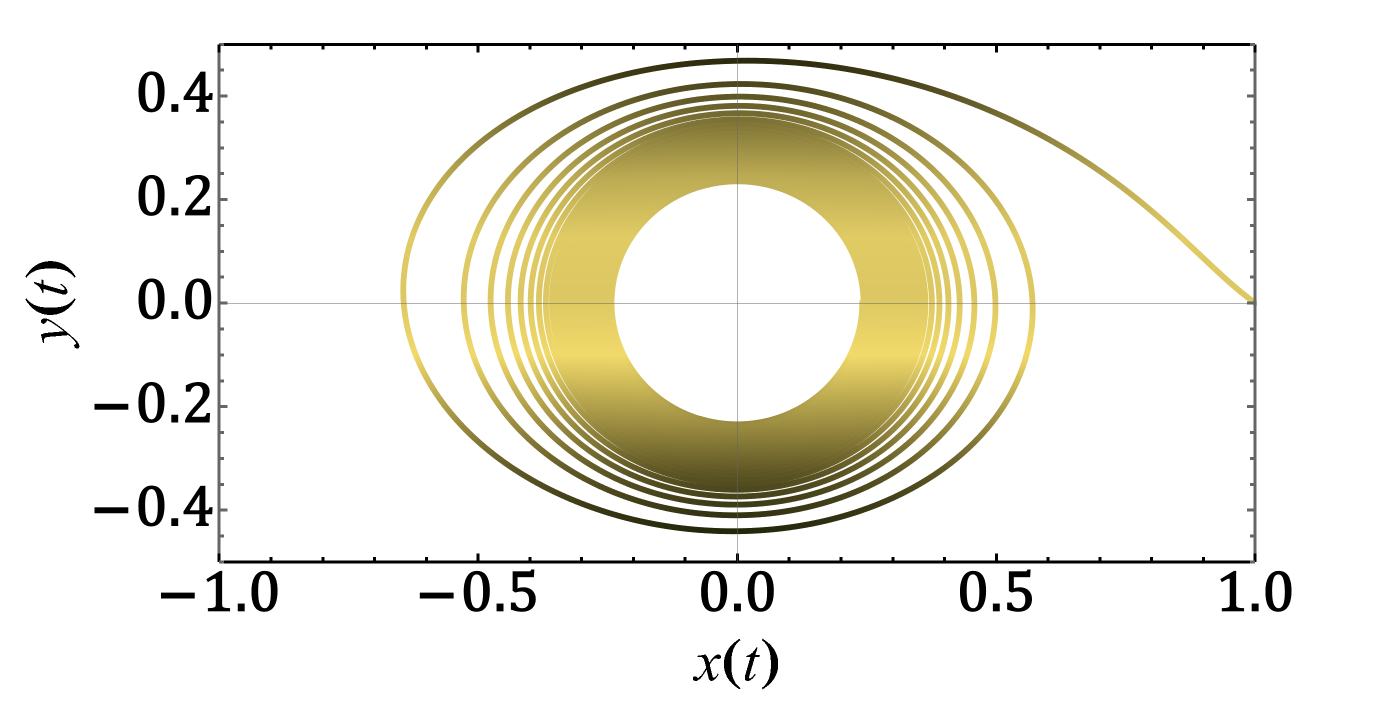}
    \caption{Orbit of the perturbed 3DIHO  under a perturbing force of the form $\boldsymbol{\delta F}=-Cv^{-3}\boldsymbol{v}$ for 32 revolutions. The initial semiaxes are defined as $a=1$ and $b=0.5$ in arbitrary units. In this case, the eccentricity decreases with time.}
    \label{Chandra}
\end{figure}

Since ${_2}F_1(3/2,1/2,2,\epsilon^2)\leq {_2}F_1(3/2,3/2,2;\epsilon^2)$ for $\epsilon^2\in[0,1)$ (as  already mentioned), the orbit evolves to a less eccentric one, as it is shown in Fig. \ref{Chandra}. As this friction is more prominent for more massive bodies, this result agrees with the known fact that massive stars tend to be found in the cluster's cores, which facilitates the runaway collision mechanism.
%
%
As the eigenvalues of $A$ are the energies of the oscillator in each axes, it is interesting to note that as $\epsilon^2\to0$ (i.e., $b\to a$), we have both $_{2}F_1\to1$ and hence $dA/dt\to -mC/(2a\omega)\mathds{1}$, which simplifies the calculation of the energy loss in the perturbed orbits in this regime.

In the following section, we show explicitly how to calculate the expression of the eccentricity for any perturbing force of the form $\boldsymbol{\delta F}_n=-m\gamma_n v^{n-1}\boldsymbol{v}$, which includes CF as a particular case.

\subsection{Generic $v-$power dragging force}

In this section, we analyze the intriguing case of dragging forces of the form 
\begin{equation}
\label{GeneralDrag}
\boldsymbol{\delta F}_n=-m\gamma_n v^{n-1}\boldsymbol{v}\, ,
\end{equation}
where $n\geq1$ and each $\gamma_n$ is a positive constant. These perturbing forces are commonly associated with air resistance. Of course, in the cosmological realm, there is no \textit{air} resistance, but this kind of force law is the most traditional one for dissipative forces and can appear, for instance,  when a star passes through a nebula cloud.

Another motivation for studying the influence of these forces in the 3DIHO is that they have already been considered in the Kepler problem, but never in the 3DIHO, as far as the authors' knowledge. In the perturbed Kepler problem, it is known that these perturbing dissipative forces do not cause any precession, but for $n>1$  the orbit becomes less eccentric with time. For the particular case where $n=1$ (linear drag),  the eccentricity of the perturbed orbit remains constant as the orbit collapses into the center of force \cite{Sivardire1984}. Hence, it would be interesting to investigate if this kind of behaviour also occurs in the perturbed 3DIHO.

\begin{figure}[H]
    \centering
    \includegraphics[width=1\linewidth]{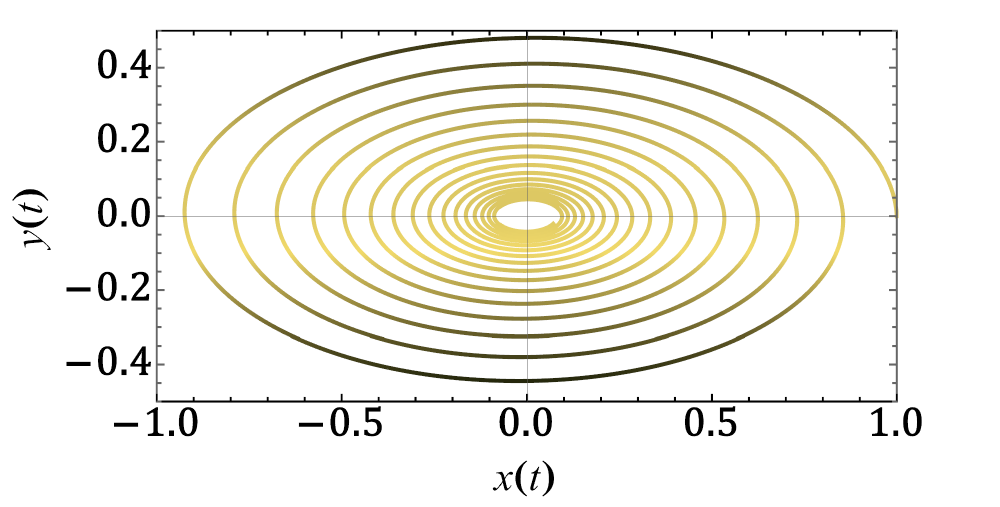}
    \caption{Orbit of the perturbed 3DIHO subjected to a perturbing force of the form $\boldsymbol{\delta F}=-m\gamma_1\boldsymbol{v}$ for 16 revolutions. The initial semiaxes are defined as $a=1$ and $b=0.5$ in arbitrary units. In this case, the eccentricity remains constant with time.}
    \label{Air1}
\end{figure}

Considering the general drag force given by Eq.(\ref{GeneralDrag}), we have for the time derivative of the bloci matrix $A$:
\begin{equation}
    \dfrac{dA^{(n)}}{dt}=-m\gamma_nv^{n-1}\begin{bmatrix}
        v_x^2 & v_xv_y\\
        v_xv_y & v_y^2
    \end{bmatrix}\, ,
\end{equation}
so that
\begin{equation}
\left\langle \dfrac{dA^{(n)}}{dt}\right\rangle=-m\gamma_n\begin{bmatrix}
    \langle v^{n-1}v_x^2 \rangle & 0\\
    0 & \langle v^{n-1}v_y^2 \rangle
\end{bmatrix}\,.
\end{equation}
For the case $n=1$, both averages become trivial,
\begin{equation}
\left\langle \dfrac{dA^{(1)}}{dt}\right\rangle=-\dfrac{m\gamma_1\omega^2}{2}\begin{bmatrix}
    a^2 & 0\\
    0 & b^2
\end{bmatrix}\,.
\end{equation}
Consequently, the perturbed block matrix $A$, denoted by $A_p$, in first approximation, can be written as
\begin{align}
A_p^{(1)}(t) &= A(0)+\left\langle \dfrac{dA^{(1)}}{dt}\right\rangle\Bigg|_0  \, t \nonumber\\&=\dfrac{m\omega^2}{2}\begin{bmatrix}
        a^2(1-\gamma_1 t) & 0\\
        0 & b^2(1-\gamma_1 t)
    \end{bmatrix}\,.
\end{align}
In this way, the eccentricity of the perturbed orbit remains  constant during the movement, but the energy diminishes with time, as the orbit  collapses towards the center of force. This result is expected when we remember that, in the traditional damped harmonic oscillator, the amplitude falls with $\mathrm{e}^{-\gamma t}$, where $\gamma$ is the damping constant. For small $t$, the energy falls with $1-\gamma t$.  In Fig. (\ref{Air1}), we plot the perturbed orbit for the case $n=1$ to show the constancy of the eccentricity during its collapse to the center of force. In this context, $\gamma_1$ can be interpreted as a time rate of relative change of the semiaxes lengths.

\begin{figure}[t]
    \centering
    \includegraphics[width=1\linewidth]{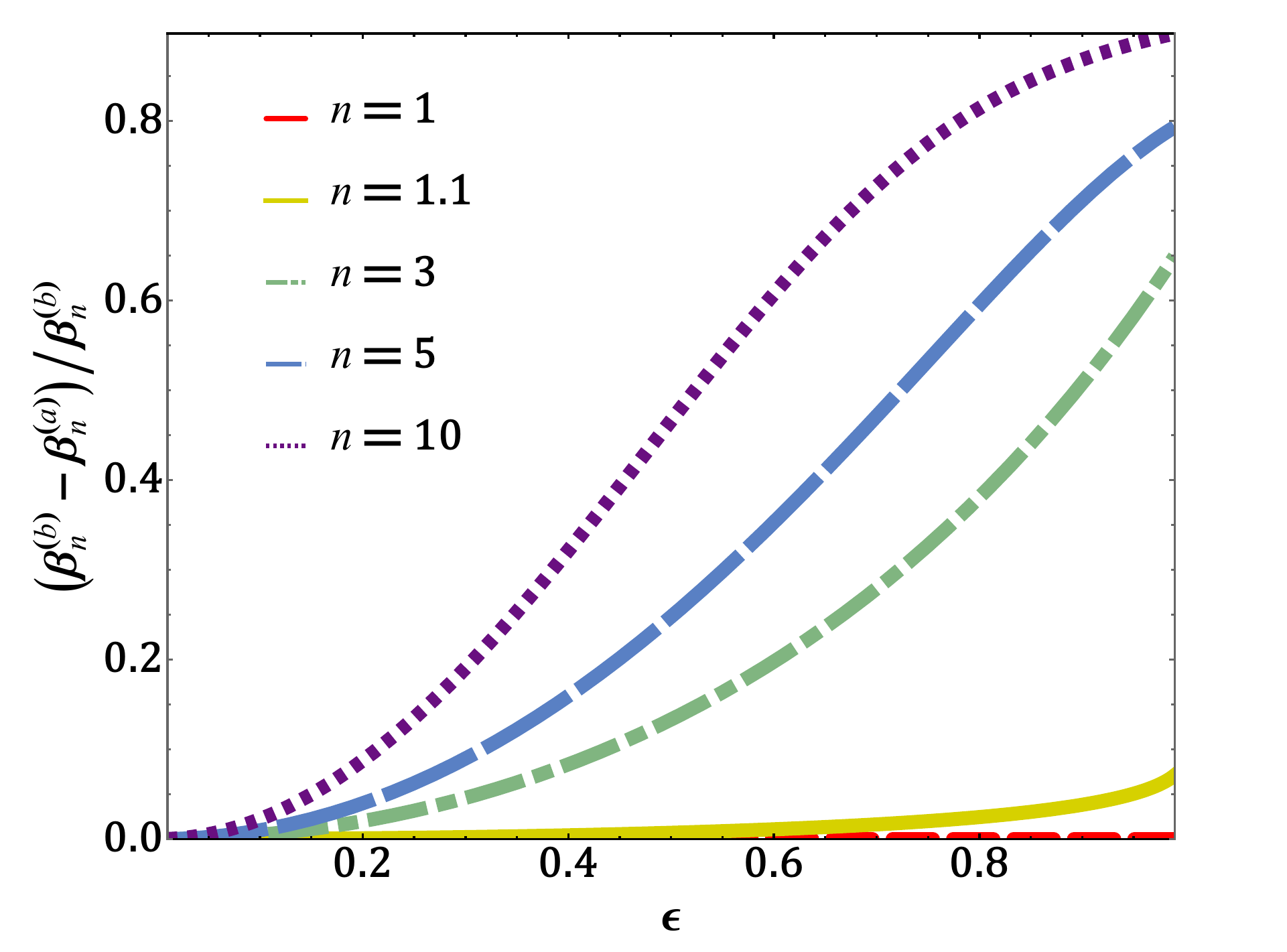}
    \caption{Relative difference between the time rates $\beta_n^{(i)}$   for $n=1,1.1,3,5,10$ as functions of the eccentricity.}
    \label{Graph}
\end{figure}

For an arbitrary $n>1$, the calculations are totally analogous to those presented in the last section, so that we just write the final results, namely, 
\begin{align}
    \langle v^{n-1}v_x^2 \rangle&=\dfrac{a^2\omega^2 (a\omega)^{n-1}}{2} {{_2}F_1([1-n]/2,3/2,2;\epsilon^2)}\,,\\
    \langle v^{n-1}v_y^2 \rangle&=\dfrac{b^2\omega^2 (a\omega)^{n-1}}{2} {{_2}F_1([1-n]/2,1/2,2;\epsilon^2)}\,.
\end{align}
As in the case of CF, the eccentricity is not  constant anymore, since
\begin{equation}
    A_p^{(n)}=\dfrac{m\omega^2}{2}\begin{bmatrix}
        a^2(1-\beta_n^{(a)} (\epsilon)t) & 0\\
        0 &b^2(1-\beta_n^{(b)}(\epsilon) t)\, ,
    \end{bmatrix}
\end{equation}
where the time rates of relative change of both semiaxes are given by
\begin{align}
\beta^{(a)}_n(\epsilon)&={\gamma_n(a\omega)^{n-1}}{{_2}F_1([1-n]/2,3/2,2;\epsilon^2)}\,,\\
\beta^{(b)}_n(\epsilon)&={\gamma_n(a\omega)^{n-1}}{{_2}F_1([1-n]/2,1/2,2;\epsilon^2)}\,.
\end{align}
Hence, the (transcendental) equation for the eccentricity of the perturbed orbit is given by\footnote{Note that this result is valid even for $n< 1$, for example, in the case of CF.}
\begin{equation}
   \epsilon=\sqrt{1-\dfrac{b^2(1-\beta_n^{(b)}(\epsilon) t)}{a^2(1-\beta_n^{(a)}(\epsilon) t)}} \,.
\end{equation}
Since ${_2}F_1([1-n]/2,1/2,2,\epsilon^2)\geq {_2}F_1([1-n]/2,3/2,2;\epsilon^2)$ for every $n\geq 1$ and for $\epsilon^2\in[0,1)$, then $\beta_n^{(b)}(\epsilon)\geq \beta_n^{(a)}(\epsilon)$ and the smallest semiaxis shrinks more rapidly than the bigger semiaxis, and so the eccentricity grows with time (in opposition with the Kepler problem). In Fig. \ref{Graph}, we plot the relative difference of the time rates $\beta_n^{(i)}$  for some values of $n$ as a function of the eccentricity. 

Note that, for a given eccentricity,  the greater the value of $n$ the greater the relative time rate.

\begin{figure}[h!]
    \centering
    \includegraphics[width=1\linewidth]{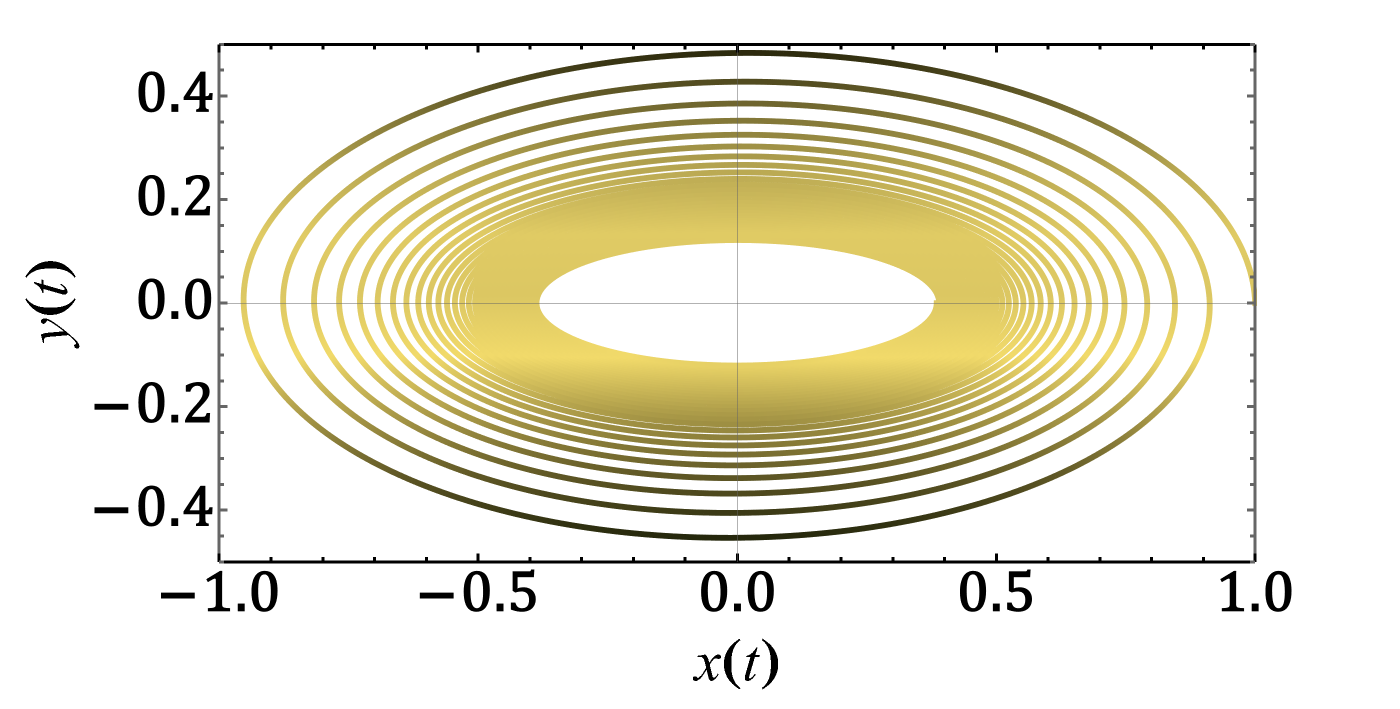}
    \caption{Orbit of the perturbed 3DIHO subjected to a perturbing drag force of the form $\boldsymbol{\delta F}=-m\beta_2v^{2}\boldsymbol{v}$ for 32 revolutions. We chose for the initial semiaxes  $a=1$ and $b=0.5$ in arbitrary units. }
    \label{Air2}
\end{figure}

\begin{figure}[h!]
    \centering
    \includegraphics[width=1\linewidth]{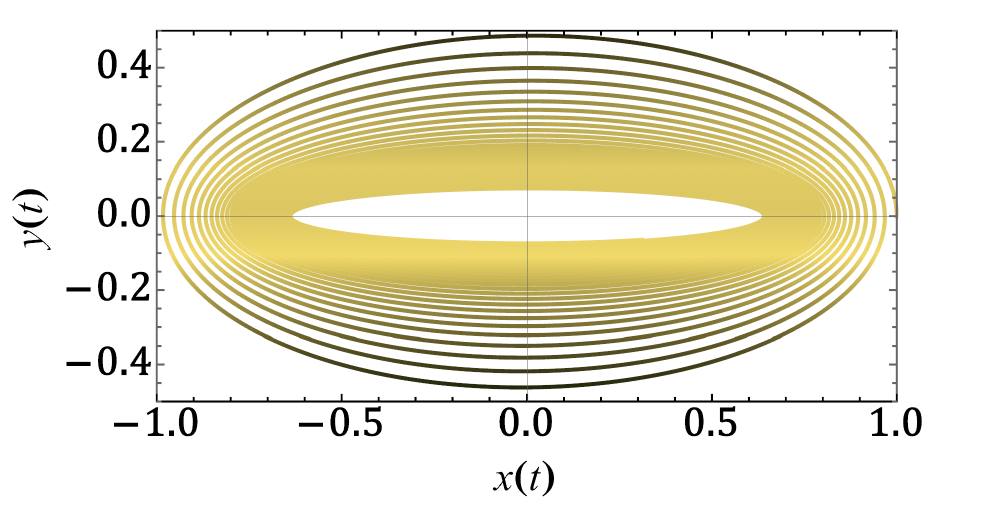}
    \caption{Orbit of the perturbed 3DIHO subjected to a perturbing drag force  of the form $\boldsymbol{\delta F}=-m\beta_5v^{5}\boldsymbol{v}$ for 32 revolutions. We chose for the initial semiaxes  $a=1$ and $b=0.5$ in arbitrary units.}
    \label{Air5}
\end{figure}

In Figs. \ref{Air2} and \ref{Air5}, we plot the orbits for $n=2$ and $n=5$, respectively, to illustrate how they become more eccentric faster as $n$ increases.

As in the case of CF, it is important to note that if $\epsilon\to0$, then $dA/dt$ will become proportional to the identity. However, if this is not the case, the orbits get more eccentric with time, and for $n>1$ there is a negative feedback: the orbit loses energy, becomes more eccentric and this gain in eccentricity makes the orbit loss more energy faster. This is immediate to see when we note that if $a,b,c\geq0$, then $_{2}F_1(a,b,c;\epsilon^2)$ is monotonically crescent with $\epsilon^2$ (see Appendix \ref{appA}). Note that this not means that the total energy is increasingly diminished, as $A_p^{(n)}$ is directly proportional to the trajectory radius, that gets smaller with time.

Comparing these results with those obtained for the perturbed Kepler problem with this same kind of perturbing drag forces, we see some similarities but also some important differences. For instance, in both cases, the perturbed orbits under this kind of dragging forces do not precess. However, an important difference relies in the fact that, concerning the time evolution of the eccentricities,  these two problems have opposite behaviours for $n>1$. While in the perturbed Kepler problem the orbits become less eccentric with time, in the perturbed 3DIHO the orbits become more eccentric with time.

\section{Conclusions and remarks}
\label{conclusion}

In this work, we analysed  the influence of different kinds of perturbing forces on the orbits of the three-dimensional isotropic harmonic oscillator (3DIHO). 
For convenience, we divided our analysis into two cases: perturbations that cause only precession, such as Larmor forces and central forces with arbitrary power laws of the distance from the star to the center of the cluster, including a keplerian-like one; and  precessionless perturbations, such as Euler forces and generic drag forces with arbitrary power laws of the velocity. We employed a method based on the Runge-Lenz invariant, a physical quantity that is conserved in the unperturbed 3DIHO, and which consists in a second rank tensor (in contrast to the Runge-Lenz vector appearing in the Kepler problem). Apart from its elegance, this method has the advantage of applying to orbits with any eccentricity.

Our analysis was primarily conducted in the context of celestial mechanics, since stars and galaxies in clusters are approximately subjected to  harmonic forces. As a first interesting application, we showed with the aid of a simple model (that can be easily generalized for more realistic situations), how to take into account that there is a concentration of mass in the core of the cluster. Concerning precessionless perturbed orbits, we highlight the so-called Chandrasekhar friction, which explains why massive stars tend to concentrate in the center of stellar clusters. We showed that, in this case, the perturbed orbit becomes less eccentric with time, a result that may be important in the analysis of the runaway collision mechanism in young and dense clusters. Finally, we would like to emphasize the differences and similarities between the perturbed 3DIHO and Kepler problems when we add a generic dragging force $\boldsymbol{\delta F}_n=-m\gamma_nv^{n-1}\boldsymbol{v}$. Although for a linear drag ($n=1$) in both scenarios the eccentricities remain constant with time, for the cases where $n>1$ we showed that in the perturbed 3DIHO the eccentricities increase with time, in contrast to what happens in the perturbed Kepler problem.



We hope this work sheds some light on perturbative methods in the study of stellar and galactic orbits inside spherical clusters. Particularly, since the orbits of stars in clusters can be used to estimate cluster masses (inverting equation \ref{Virial}), these corrections may also be useful to refine such mass estimates.
%
%

\section*{Acknowledgments and Funding}
\label{ack}

The authors are thankful to Reinaldo F. de Melo e Souza,  Ribamar R.R. Reis, I.B. Batista and W.J.M. Kort-kamp for the insightful discussions. J. O.-C thanks the Brazilian agency FAPERJ (masters scholarship No. 201.879/2025). C.F. thanks the Brazilian agencies CNPq 
(Grants No. 308641/2022-1 and 408735/2023-6) and FAPERJ (Grant No. 204.376/2024).

\appendix

\section{\label{appB}Other Approach to the Runge-Lenz Invariant for the Harmonic Oscillator}

It is possible to define an analogue to the second rank tensor $\mathds{A}-$, but using vectors, in a similar way to what is usually done in the Kepler problem, in which the Runge-Lenz vector is defined as
\begin{equation}
    \boldsymbol{A}=\boldsymbol{p}\times\boldsymbol{L}-mr^2\boldsymbol{F}\,,
\end{equation}
 where $\boldsymbol{L}=\boldsymbol{r}\times \boldsymbol{p}$ is the angular momentum of the particle with respect to the center of force and $\boldsymbol{F}$ is the total force acting on the particle. In order to create an invariant similar to this vector, we start by writing the eigenvectors of $\mathds{A}$, which are given by \cite{Sivardire1989}
\begin{align}
\boldsymbol{A}^{(x)}&=\boldsymbol{p}\times\boldsymbol{L}-ma^2(m\omega^2)\boldsymbol{r}\,,\\
\boldsymbol{B}^{(y)}&=\boldsymbol{p}\times\boldsymbol{L}-mb^2(m\omega^2)\boldsymbol{r}\,.
\end{align}
Note that these vectors have the same form as the Runge-Lenz vector in Kepler's problem, but although their directions remain constant with time their modula are not conserved. Hence, in order to obtain vectors which are a conserved quantities in the HO problem, all we need to do is to divide both vectors by their modula:
%
%
\begin{align}
\boldsymbol{a}^{(x)}&=\dfrac{\boldsymbol{p}\times\boldsymbol{L}-ma^2(m\omega^2)\boldsymbol{r}}{\omega^2\sqrt{(a^2-b^2)(r^2-b^2)}}\,,\\
\boldsymbol{b}^{(y)}&=\dfrac{\boldsymbol{p}\times\boldsymbol{L}-mb^2(m\omega^2)\boldsymbol{r}}{\omega^2\sqrt{(a^2-b^2)(a^2-r^2)}}\,.
\end{align}
Note that if we use the coordinate system adopted in this paper, $\boldsymbol{a}^{(x)}$ and $\boldsymbol{b}^{(y)}$ will be just a complicated way of writing $\hat{x}$ and $\hat{y}$ in terms of $m,\omega,\boldsymbol{r}$ for generic $a,b$.

It is worth emphasizing that these invariant vectors can be used to obtain the orbit equation in the same fashion as what it is  done in the Kepler problem with the Runge-Lenz vector. 
Besides, the results obtained in this paper can also be re-derived 
working directly with these vectors (instead of using the $\mathds{A}-$invariant), but the calculations may become much more involved. 
%
%
If for a given perturbing force one obtains, for example, $\dfrac{d\boldsymbol{a}^{(x)}}{dt}=\boldsymbol{\Omega}\times \boldsymbol{a}^{(x)}$, then $\Omega$ will be readly identified as the velocity of angular precession.

\section{\label{appA}The Hypergeometric Functions}

Throughout this paper, we made use of the so-called hypergeometric function ${_2}F_1(a,b,c;z)$. It is not a surprise that this function appears in the study of orbits, since it is a generalization of the elliptic functions. The hypergeometric function is the solution of Euler's (hypergeometric) differential equation \cite{andrews1999},
\begin{equation}
    z(1-z)\dfrac{d^2{_2}F_{1}}{dz^2}+(c-(a+b+1)z)\dfrac{d{_2}F_{1}}{dz}-ab{_2}F_{1}=0\,.
\end{equation}
The first time that this function appeared in the literature was in the context of a hypergeometric series in John Wallis book \textit{Arithmetica Infinitorum} \cite{Wallis}, in 1655. In modern notation, for $|z|<1$ and $c\notin\mathds{Z}_-$, the hypergeometric function is given by the power series
\begin{align}
{_2}F_1(a,b,c;z)&=\sum_{n=0}^{\infty} \frac{(a)_n (b)_n}{(c)_n} \frac{z^n}{n!} \\&= 1 + \frac{ab}{c} \frac{z}{1!} + \frac{a(a+1)b(b+1)}{c(c+1)} \frac{z^2}{2!} + \cdots\nonumber \,,
\label{PowerSeries}
\end{align}
where the second equality is valid if $a,b,c\in\mathbb{Z}$ and with
 $(p)_n$ being the  Pochhammer symbol,
\begin{equation}
    (p)_n=\dfrac{\Gamma(p+n)}{\Gamma(p)}\,.
\end{equation}
From the above expression, it is obvious that ${_2}F_1(3/2,3/2,2,\epsilon^2)\geq {_2}F_1(3/2,1/2,2,\epsilon^2)$, as it was stated in Secs. \ref{sec3} and \ref{sec4}. It is also immediate to see that $_{2}F_1(0,b,c;z)=1$, which  together with Eq.( \ref{GeneralPrecession}), shows that for a harmonic perturbation, there will be no precession. Finally, note that if $a,b,c\geq0$, then $_{2}F_1(a,b,c;z)$ will   increase monotonically with $z$ as we stated in Sec. \ref{sec4}. Besides the qualitative benefits of the above definition for $_{2}F_1$, a nice use of this function in calculations is related to the Gauss integral formulation, which is always valid if $\Re\{z\}<1$,
\begin{equation}
    _2F_1(a,b,c;z)=\dfrac{\Gamma(c)}{\Gamma(b)\Gamma(c-b)}\int_0^1 t^{b-1}(1-t)^{c-b-1}(1-zt)^{-a}dt\,.
\end{equation}

To get the expressions we used in this paper, we can perform the variable transformation $t=\sin^2(x)$, and then convert the previous equation into the following one:
\begin{widetext}
\begin{equation}
    _2F_1(a,b,c;z)=\dfrac{2\Gamma(c)}{\Gamma(b)\Gamma(c-b)}\int_0^{\pi/2} \sin(x)^{2b-1}\cos(x)^{2c-2b-1}(1-z \sin^2(x))^{-a}dx\,.
\end{equation}

Analogously, if we choose $t=\cos^2(x)$, one gets
\begin{equation}
    _2F_1(a,b,c;z)=\dfrac{2\Gamma(c)}{\Gamma(b)\Gamma(c-b)}\int_{0}^{\pi/2} \cos(x)^{2b-1}\sin(x)^{2c-2b-1}(1-z\cos^2(x))^{-a}dx\,.
\end{equation}
\end{widetext}

In this paper, we used only the following values of the Gamma function:  $\Gamma(2)=1$, $\Gamma(1/2)=\sqrt{\pi}$, and $\Gamma(3/2)=\sqrt{\pi}/2$.
Various special functions can be written as particular cases of the hypergeometric one. For example, both complete elliptic integrals are such that
\begin{align}
K(\epsilon)&=\dfrac{\pi}{2}{_2}F_{1}(1/2,1/2,1,\epsilon^2) \quad\text{and}\\ E(\epsilon)&=\dfrac{\pi}{2}{_2}F_{1}(-1/2,1/2,1,\epsilon^2)\,.
\end{align}

\bibliographystyle{ieeetr}
\bibliography{apssamp}

\end{document}